# Bragg-like curve for dark matter searches; binary gases

Akira Hitachi
*Molecular Biophysics, Kochi Medical School, Nankoku, Kochi 783-8505, Japan*
*E-mail:* jm-hitachia@kochi-u.ac.jp

**Abstract**
Bragg-like curves, the electronic energy deposition as a function of the projected range, for recoil ions in $N_2$, $CO_2$, $CF_4$ and $CS_2$ are presented. The curves are intended for directional search for the dark matter candidate, WIMPs, using the gas TPC. Nuclear quenching factors for very heavy ions produced in α-decay in dry air, $N_2$, $CO_2$, $CF_4$ and $CS_2$ are also estimated and compared with experimental values.

## 1. Introduction

Most of the constituent matter of the universe is dark with no emission, no absorption, even no scattering of light. It is, in fact, transparent. In 1933, Zwicky announced that there is a missing mass in the universe by observing the rotational velocity of galaxies in the Coma cluster. Further scientific evidence confirmed his observation: rotational velocity in a galaxy, gravitational lensing, the microwave radiation background, even the shape of thin galaxies require unseen mass to keep a galaxy from collapsing (IDM2006; Sumner 2002). The unseen dark matter accounts for a quarter of the universe. Ordinary matter makes up only 4%. The rest, ~73%, is supposed to be dark energy.

In recent years, researchers have gone underground to look for WIMPs (weakly interacting massive particles), the leading candidates for dark matter. They are searching for recoil nuclei of a few tens of keV energy, produced by elastic scattering with WIMPs. However, almost all the signals produced in an underground detector are from the background, mostly γ-rays. It is essential to distinguish the WIMP signal from the background. Here, the radiation physicist and chemist can help to predict the footprints of WIMPs and the like.

One of the methods to identify WIMPs would be the directionality of the recoil ions (Spergel, 1988; CYGNUS2007). The solar system is traveling around the galactic center at about 230 km/sec. The Earth's orbit around the Sun is inclined at an angle of 60º to the galactic plane and the Earth's spin axis is 23.5º away from its orbit. Then, the direction of a "WIMP window" (or the flux of WIMPs) would change annually or even daily. Most of the detectors can observe the annual modulation. However, only detectors with directional capabilities, such as the TPC (time projection chamber), can observe daily fluctuations. Many new technologies for constructing micro-size-ion tracks have been developed. Typical gases used for these TPCs are electronegative low diffusion gases such as $CS_2$ (Snowden-Ifft et al., 2000). Another gas often used is $CF_4$ in which electrons are the charge carriers. $CF_4$ is chosen because fluorine is considered to be very effective for WIMP search via spin-dependent interaction (Ellis and Flores, 1991).

The nuclear stopping $S_n$ is of the same order of magnitude as that of the electronic stopping $S_e$ for the interaction of ions of a few keV energy with matter (Lindhard et al., 1963). The secondary ions will go into the collisional processes again, and so on. After cascade processes of stopping collisions, considerable amounts of energy go into atomic motion which is wasted as heat in ordinary detectors. Only a part of the energy η goes to the electronic excitation which can be observed as ionization or scintillation signals. The nuclear quenching factor, or the Lindhard factor, is expressed as $q_{nc} = \eta/E$. The quenching factors are defined as the energy ratio that escapes quenching, i.e. $q_{nc} = 1$ for no quenching and $q_{nc} = 0$ for total quenching.

The linear energy transfer (LET) is a major factor in understanding radiation effects. The LET is simply $-dE/dx$ for fast ions, since the total energy loss $S_T$ is due almost exclusively to electronic stopping; $S_T \approx S_e$. However, it is necessary to introduce the electronic LET ($LET_{el} = -d\eta/dx$) for very slow ions since $S_T = S_n + S_e$ (Hitachi, 2005, 2007). In other words, $LET_{el}$ is the specific electronic energy loss along the track of charged particle. A particle of energy $E_0$ can deposit an electronic energy $\eta_0$ in a range $R_0$. The particle energy becomes $E_1$ after an energy loss of $\Delta E$, then it can deposit $\eta_1$ in $R_1$. $LET_{el}$ can be defined as the electronic energy deposited per unit length travelled by the particle.

$$LET_{el} = - d(qE)/dR = - d\eta/dR \approx -\Delta\eta/\Delta R = - (\eta_1 - \eta_0)/(R_1 - R_0) \qquad (1)$$

$LET_{el}$ for slow ions can be compared with LET for fast ions. Then, the knowledge accumulated over many years in radiation physics and chemistry can be used to obtain the scintillation and ionization properties in condensed media such as recoil Xe ions in liquid Xe (Hitachi, 2005).

The evaluation of the range-ionization curve is required for the directional detection of recoil ions in a gaseous TPC. The Bragg curve, the specific energy loss, $-dE/dx$, along the track of a charged particle, gives a range-ionization curve for the fast ions such as α-particles. For fast ions, the track is almost straight and the $w$-value, the average energy required to produce an ion pair, can be regarded as constant. However, for slow recoil ions, the total energy deposited is different from the electronic energy deposited as discussed above, and the trajectories are tortuous and ramify. Similar to the $LET_{el}$, we introduce the Bragg-like curve, $-d\eta/dR_{PRJ}$, which is closely related to the range-ionization curve, using the projected range (the distance of penetration), $R_{PRJ}$ in Eq. (1) (Hitachi 2007). The curve can give an averaged, one dimensional ionization distribution of the recoil ion track from head to tail. The Bragg-like curve was introduced for practical purposes in gas TPC. The values of $S_T$, $S_n$, $S_e$, and $LET_{el}$, together with the Bragg-like curve, for N ions in nitrogen are compared as a function of energy in Fig. 1. $LET_{el}$ is larger than $S_e$ but smaller then $S_T$. The values for $S_T$, $S_e$, and $LET_{el}$ become the same for fast ions.

## 2. Recoil ions in binary gases

### *2.1. The nuclear quenching factor*

Lindhard et al. (1963) solved the homogeneous integral equation for ν (= ε - η) in a single-element material and gave numerical results for the nuclear quenching factors $q_{nc} = \eta / E$ for $Z_1 = Z_2$ and for $k$ = 0.1, 0.15 and 0.2. Here $Z_1$ and $Z_2$ are the atomic numbers for the projectile and the target, and $k$ is a parameter associated with the electronic stopping power, $(d\varepsilon/d\rho)_e = k \cdot \varepsilon^{1/2}$ (ε and ρ are the dimensionless reduced energy and range, respectively). The model has explained the recoil ion to γ, RN/γ, ratios in semiconductors (Gerbier 1990) and the ionization measurement for Ar ions in Ar gas (Madsen 1945; Hitachi 2007).

On the other hand, the formulation of the general solutions for $q_{nc}$ becomes quite complicated where the medium contains more than one element and further approximations are necessary in binary gases. The power law approximation, which will be discussed later, was introduced by Lindhard et al. for very heavy ions, but it does not apply to light elements. We will attempt two methods for recoil ions in binary gases: (a) the asymptotic form with $Z_1 \neq Z_2$ and (b) an independent element approach with $Z_1 = Z_2$.

An asymptotic form was introduced by Lindhard et al. (1963) for $Z_1 = Z_2$ with $0.1 < k < 0.2$. It reads

$$\nu = \frac{\varepsilon}{1 + k \cdot g(\varepsilon)} \qquad (2)$$

The function $g(\varepsilon)$ was originally given by a figure and was recently fitted by Lewin and Smith (1996). The asymptotic form reproduces the numerical ν within an accuracy of several %. The form was tried here for $Z_1 \neq Z_2$ in binary gases. Also, in the approximation, the target molecule was replaced by an element which has $Z$ and $A$ values close to averaged $Z$ and $A$ of the molecule. Namely, $CS_2$ molecules were replaced by Al and $q_{nc}$ values for C ions in Al and S ions in Al were obtained by using Eq. (2). Since the average for $Z$ and $A$ for $CF_4$ are quite close to those of F, the target molecule $CF_4$ was simply replaced by F and $q_{nc}$ values for C ions and F and F ions in C were calculated. This has also been applied to $CO_2$.

The independent element approach was also taken to obtain $q_{nc}$ in binary gases. $CX_m$ may be treated by taking C ions in C and X ions in X, instead of C ions in $CX_m$ and X ions in $CX_m$, respectively. Then, the Lindhard factors $q_{nc}$ are calculated using the numerical values for $k = 0.15$, or Eq. (2) when $k$ is not close to 0.10, 0.15 or 0.20.

*2.2. The Bragg-like curve*

The electron is taken as the charge signal because of its high velocity in most gas detectors. However, its high mobility may be a disadvantage because of its large diffusion in the position determination in TPC for very short ion tracks. Alternatively, electronegative molecules may be used in a TPC for WIMP searches. An ionizing particle enters TPCs filled with electronegative low diffusion gases, such as $CS_2$, produces many electrons, ions and fragments. The ejected electron immediately attaches to a surrounding neutral molecule and produces a negative ion. The negative ion will drift towards a two-dimensional-position-sensitive electrode (x-y). The negative ion releases the electron under a high electric field near the anode. The freed electron undergoes the usual electron multiplication process. The time difference between the incidence of the projectile and the arrival times of negative ions give the third dimension (z). The spatial distribution of the charge produced by the projected ion can be reconstructed to give the directionality of WIMPs in a TPC. We calculate here the Bragg-like curve, $d\eta/dR_{PRJ}$, for a one dimensional ionization distribution as a function of projectile direction $R_{PRJ}$.

The projected range, $R_{PRJ}$, was obtained from SRIM2006. The stopping power of the compound is obtained using the Bragg rule, in which the stopping power of a compound is simply given by the linear combination of the stopping powers of the individual elements. However, the outer shell electrons have different orbitals in the compound than in the corresponding elements and the core and bond approximation is often taken (Ziegler and Manoyan, 1988). The compound correction may be needed for molecules containing only the light elements, such as H, C, N, F, O, in $CO_2$, $CF_4$ and $N_2$. However no correction was applied here because it is not accurate for $Z_1 > 3$ (SRIM2006). The correction for $CS_2$ is not needed since $CS_2$ contains a heavy atom, S.

## 3. Heavy recoil ions in α-decay

In α-decay, the daughter nucleus recoils as a very heavy ion with typically 100-200 keV. The very heavy recoil ions in α-decay produce WIMP-like signals in detector media, and their contribution to the background signal can be very serious. For example, $^{222}$Ra decays to produce Po (or Pb) ions and α recoils. Usually, one does not see these as separate particles, because they are produced at the same time. The Po signal is associated with the much larger α signal. However, in some systems Po decays such that an α particle is stopped in the wire or wall and the Pb recoil fully goes into the gas. Then, the recoil Pb ion can produce a WIMP-like signal. It is important to know what signal will be produced.

We use a power law approximation for $Z_1 \neq Z_2$ at very low energy (Lindhard et al., 1963),

$$\eta = CE^{3/2}, \quad \text{for } E < E_{1c}, E_{2c} \tag{3}$$

where $C = ⅔\{E_{1c}^{-1/2} + ½\gamma^{1/2}E_c^{-1/2}\}$, $\gamma = 4A_1 A_2 /( A_1 + A_2 )^2$ and $E_c = \gamma E_{2c}$. Two characteristic energies, $E_{1c}$ and $E_{2c}$ associated with $Z$ and $A$ for the target atom and the projectile atom, set the upper boundary. The energy $E_{2c}$ sets the lower criterion when $Z_1 > Z_2$ as in the cases here.

The values of $q_{nc}$ for Pb ions in α-decays in compounds are obtained using the power law approximation additive property; $q_{nc}$ for a compound is given by the linear combination of $q_{nc}$ of individual elements, e.g., $q_{nc}$ ($CS_2$) = [$q_{nc}$ (C) + 2 $q_{nc}$ (S)]/3. Thereby the values for $H_2$, $CF_4$, $CO_2$ and dry air (4N + O) are also calculated.

## 4. Result and discussion

Stopping powers (Biersack et al., 1975), electronic LET and Bragg-like curves for N ions in nitrogen are shown as a function of the ion energy in Fig. 1. The Bragg peak is at about 6 MeV and far right, outside the figure. The $LET_{el} = -d\eta/dR_T$, where $R_T$ is the true range, is larger than the electronic stopping power $S_e$, particularly in the low energy region where the contribution from the nuclear stopping power $S_n$ becomes comparable to or larger than $S_e$. The Bragg-like curve, $-d\eta/dR_{PRJ}$, is larger than $LET_{el}$, because the ion track at low energy is tortuous and have some branches, $R_{PRJ}$ becomes shorter than $R_T$. At high energy, the contribution from the nuclear stopping becomes negligible and the curves for $S_T$, $S_e$, $LET_{el}$, and $d\eta/dR_{PRJ}$ become practically the same for fast ions such as protons, alphas and even for fission fragments. The Bragg-like curve for N ion in $N_2$ is also plotted as a function of $R_{PRJ}$ in Fig. 2. The N ions enter from the right hand side. The energy η shown here is only the electronic energy that can be used for scintillation or ionization. The energies of ions recoiled by WIMPs are not monochromatic but have a distribution. The projectile energy starts at 200 keV in the figure, however the same curve applies for any energy of 10 keV to 200 keV. The area below the curve (with the unit shown on the axis on the right) expresses the number of ions produced, $N_i = \eta/W$, with a $W$-value of 36.6 eV for $N_2$.

### 4.1. *Recoil nuclei*

The $q_{nc}$ values obtained for the recoil ions in $CO_2$, $CF_4$ and $CS_2$ are shown in Figs. 3-5. The independent element approach ($Z_1 = Z_2$, solid curves) and the asymptotic form ($Z_1 \neq Z_2$, dot-dashed curves) give practically the same values at high energy for all the gases. The difference between the C ion and the other ions (O, F or S) was wider in the asymptotic form than in the independent element approach. The difference is larger at low energies.

The electronic to total stopping power ratio, $S_e/S_T$ are similar for C ions in C and C ions in S. This is also true for S ions in S and S ions in C. Therefore the simple estimate taken here for $q_{nc}$ may be valid in $CS_2$. The $q_{nc}$ values estimated for recoil C and S ions in $CS_2$ are shown in Fig. 3. The values calculated for S ions in Al and in S by the asymptotic form are quite close; 0.28 and 0.29 at 10 keV and 0.57 and 0.57 at 200 keV, respectively. Therefore, the results for S ions in C is shown in Fig. 3. Similarly, the results for C ions in Al and in S are very close, therefore the results for C ions in S is shown for the asymptotic form in Fig. 3, as also the results shown for other gases in Figs. 4-5. The $q_{nc}$ values calculated for $CS_2$ are compared with the experimental work of Snowden-Ifft et al. (2003) which are not pure experimental values but in part simulation. The independent element approach (S ions in S) agrees quite well with the measured values for S ions in $CS_2$, however, the values calculated for C ions (C in C) give higher values at > 100 keV. The asymptotic form for S ions in C agrees well at high energy but gives a smaller value at low energy. The asymptotic form for C ions in S agrees quite well at low energy but it gives larger values at high energy. The reason why the experimental values for C ions are lower at high energy is not well known.

The values obtained by the power law approximation, C in Al and S in Al, are also shown with broken curves for comparison. The energy criteria set by $E_{1c}$ or $E_{2c}$ are 24 keV and 177 keV, respectively, for C and S ions in Al. The power law approximation gives quite steep curves and the values are large except at very low energy. The energy criterion may be lower than $E_{1c}$ or $E_{2c}$ for light ions.

Fig. 4 shows the results for $CF_4$. No experimental values for $q_{nc}$ are available for $CF_4$. The asymptotic form ($Z_1 \neq Z_2$) gives larger values for C ion and smaller values for F ions than the independent approach. However, the differences are small, and both approximations give practically the same values for C and F ions in $CF_4$ except at a very low energy. This may be due to the $Z$ and $A$ values for C and F atoms being not very different. Similar results were obtained for C and O recoil ions in $CO_2$ as shown in Fig. 5. The $q_{nc}$ values for N ions in $N_2$ are also shown in Fig. 5. The values lay between the curves for C and O ions as one might expect.

The Bragg-like curves calculated for recoil ions from 10 to 200 keV in $CO_2$, $CF_4$ and $CS_2$ are shown in Figs. 6-8. C ions are shown with closed circles and the partner ions are closed squares. The value of $q_{nc}$ obtained by the independent element approach was taken. The points are plotted in every 5 keV for 10 to 50 keV, in 10 keV for 50 to 100 keV, and in 20 keV for 100 to 200 keV. $W$-values of 33 eV, 54 eV and 19 eV were used for $CO_2$, $CF_4$ and $CS_2$, respectively, to obtain $\Delta N_i$. The area below each curve (read with the axis on the right) expresses the number of ions produced, $N_i = \eta/W$.

The Bragg-like curves for C and S ions in $CS_2$ show strong dependences on energy (Fig. 6) and look suitable for the directional detection of the WIMPs. However, the curves for C and F ions in $CF_4$ do not have strong energy dependences except below 50 keV. The curve for F ions, which is expected to have a favourable spin dependent interaction, is nearly flat above 50 keV. The number of ions produced in $CF_4$ is also much smaller than that for $CS_2$ because of the large *W*-value (54 eV). The curves for C and O ions in $CO_2$ are between those for $CS_2$ and $CF_4$.

The incident ion can produce high energy recoils of other elements which can ionize the target molecules. For example, the incident C ion may produce S ions in $CS_2$, and the S ions can contribute to the ionization. This effect may not be included in the independent element approach. It is assumed that the ionization is due to the projectile element in the present calculations for $q_{nc}$ for recoil atoms. It is not clear how much of such effect is included in the asymptotic form. Ling and Knipp (1950) presented a model for Pb ions in Ar similar to the power low approximation by Lindhard. They estimated that about 2/3 of the ionization is due to the secondary ions. If the contribution from the other element ion is large, the curves for C ions and S ions in $CS_2$ in Fig. 3 may shift nearer to each other. This may explain the behaviour of the experimental $q_{nc}$ curve for C ions. However, it is still strange that the deviation occurs at high energy. The contribution from scattering is smaller at high energy.

### 4.2. *Heavy recoil ions in α-decay*

The α-decay from $^{210}$Po, ThC ($^{212}$Bi), and ThC' ($^{212}$Po) produces recoil ions of 103 keV $^{206}$Pb, 117 keV $^{208}$Tl and 168 keV $^{208}$Pb, respectively. Also, $^{212}$Rn daughters, $^{218}$Po gives 112 keV $^{214}$Pb, and $^{214}$Po gives 147 keV $^{210}$Pb. Nuclear quenching factors $q_{nc}$ for heavy recoil ions in α-decay of Po, ThC and ThC' in dry air, $N_2$, $CO_2$, $CF_4$ and $CS_2$ are obtained by the power law approximation and are shown in Table I. The ionization measurements in a gas basically give the Lindhard factor $q_{nc}$ both for the gaseous and condensed phases: $q_{nc} = W(\alpha)/W(\mathrm{RN})$, where $W(\alpha)$ and $W(\mathrm{RN})$ are the *W*-values for fast ions and slow recoil ions, respectively. Calculated $q_{nc}$ values are compared with $W(\alpha)/W(\mathrm{RN})$ values measured in gas phase. The values calculated for $N_2$, $CO_2$, $CF_4$ and $CS_2$ agree well with the experimental values (Stone and Cochran, 1957, Cano, 1968). The general trend is that $q_{nc}$ decreases as the target $Z_2$ increases.

The power law approximation does not apply for $H_2$ because $E_{2c}$ becomes too small but it is shown just for comparison. The calculated $q_{nc}$ values are much larger than the measured values. Hydrocarbons have a strange behaviour. The measured values of $q_{nc}$ are 0.457and 0.543 respectively for 117 keV and 168 keV in $H_2$ and these are still much higher than those for C, O, F and S. If the general trends with $Z_2$ apply also for the hydrocarbons, $q_{nc}$ values for $CH_4$, $C_2H_4$ and $C_3H_6$ should be higher than those for $CO_2$, $N_2$ and dry air. However, the reverse is observed.

The values obtained by the power law approximation for heavy recoil ions, Pb, in α decay (dashed curve) are also shown for *ca* 100 keV to 170 keV in Figs. 3-5 with dashed curves. The thin dot-dot dash curves, e.g., Pb ions in C and Pb ions in S, and Pb ions in Al, are shown in Fig. 3.

The Bragg-like curves for Pb ions in $N_2$, $CO_2$, $CF_4$ and $CS_2$ are obtained using the $q_{nc}$ values given by the same method as recoil ions and are shown in Fig. 9. The ion enters from the right. They give roughly straight lines. This may be because the same energy dependence, $q_{nc} \propto E^{1/2}$, was assumed in the power law approximation. The ionization densities for very heavy recoil ions in α-decay is much higher than those for recoil ions and the range is much shorter. The total amount of the ionization produced by heavy recoil ions, shown as the area below the Bragg-like curve, is in the same order of magnitude as those for recoil ions. Therefore, the very heavy recoil ions in α-decay may give WIMP-like signals in detectors without track shape, or length, discrimination capabilities.

4.3 *General remarks*

Compounds containing only light elements, hydrocarbons and $CF_4$, may require careful treatment, since the compound corrections for the stopping power of these compounds are required. The Bragg rule introduces errors of several percent in the stopping power at low energies because the electronic wavefunctions of the target depends on the bonds of each atom with its neighbours. This effect is most important for the low-$Z$ atoms of biological interests (Hobbie, 1987). The $W$-values may also have considerable energy dependence for these compounds (Combecher, 1980). In addition to this, the triplet states in these molecules are metastable which are not excited by fast particles but can be excited effectively by slow heavy ions. These facts may affect the stopping powers, $q_{nc}$, and $W$-values.

The validity of the asymptotic form for $Z_1 \neq Z_2$ with $0.1 < k < 0.2$ is still in question. It looks as if the form is still good for recoil ions in binary gases as shown for $CO_2$, $CF_4$ and $CS_2$. The values for $k$ are within 0.10 to 0.20 for recoil ions in those gases except for $CS_2$. However, when the projectile becomes very heavy the difference in $Z_1$ and $Z_2$ becomes large, then the asymptotic form seems to fail. The values obtained for heavy recoil ions in α-decay are small in most gases. The $k$ values are 0.10 to 0.16 and are generally in the range of 0.1 - 0.2. The $q_{nc}$ values calculated using the asymptotic form for 103 keV Pb ions are 0.20 - 0.21 in He, C, N, O, and S, and those for 168 keV Pb ions are 0.22 – 0.23 and are practically the same, whereas reported experimental values are scattered as shown in Table I. The discrepancy may be due to the fact that the asymptotic form does not include the effects of the secondary ions.

A cylindrical picture of the ion track may be obtained by assuming a Gaussian in the lateral direction and taking lateral straggling ($\Delta R_\perp$). It is hard to take the longitudinal straggling ($\Delta R_{/\!/}$) into account but it is possible. Such a two-dimensional picture with cylindrical geometry represents an ensemble average of many ions. However, the shape of each ion track changes because of the scattering. Additionally the amount of ionization produced is not large. The individual track may stay within the bell-shape contour (with $\Delta R_\perp$) or the drop-shape contour (with $\Delta R_\perp$ and $\Delta R_{/\!/}$). However, the ionization may be distorted and localized. The effect of the straggling on each track is not well known.

Evans et al. (1963) observed ionization density contours for He, N, Ne, and Ar ions in He, $N_2$, air and Ar gases in the range from *ca* 20 keV to 250 keV. They reported that the attenuation of the beam of ions is found to be approximately exponential in the axial direction and approximately Gaussian in the lateral direction. The Bragg-like curve obtained for N ions in nitrogen compare well with the measured range-ionization curve for 57 keV N ions in air except at the very beginning. However, the range-ionization curve shows less ionization near the end of the track. This may be due to the fact that the present calculation does not take longitudinal straggling into account. The values for 57 keV N ions in $N_2$ are $R_{PRJ} = 3.25$, $\Delta R_{/\!/} = 0.89$, and $\Delta R_\perp = 0.75$, $\times 10^{-3}$ mg/cm$^2$ (SRIM). The charge distribution spreads in the longitudinal as well as in the lateral direction. The longitudinal spread reduces the charge density particularly at the end in the range-ionization curve. This is encouraging for the directional detection of WIMP foot prints. The directionality may be clearer in actual range-ionization curve than the Bragg-like curve without the straggling as presented here.

The Bragg-like curve becomes steeper for heavy elements. It may be useful to use gas mixtures such as Ar + M or Xe + M, where M is a electro-negative molecule, in TPC for dark matter searches as far as the directionality is concerned. However, at very low energy, the scattering becomes large for heavy atoms.

## 5. Summary

The Bragg-like curves, which are closely related to the range-ionization curve, for recoil ions in $N_2$, $CO_2$, $CF_4$ and $CS_2$ were obtained for the directional detection of recoil ions in gas TPC. The present results show that the information on the change in the nuclear quenching factor, $q_{nc}$, as a function of the energy is not enough to consider directional capabilities of dark matter detectors. One needs the Bragg-like curve which includes changes in $q_{nc}$ and the stopping powers. The power law approximation for $Z_1 \neq Z_2$ gave satisfactory values of $q_{nc}$ for the heavy recoil ions in $\alpha$-decay except in the hydrocarbons.

## Acknowledgements

The author would like to thank Dr. A. Mozumder and Dr. J. A. LaVerne for helpful discussions. He also thanks Dr. P. Majewski and Dr. K. Miuchi for useful information. He is also grateful to Prof. T.A. King for reading of the manuscript.

Table I

Nuclear quenching factors $q_{nc}$ for recoil ions in α-decay of Po, ThC and ThC' in various gases. The experimental values are obtained assuming $q_{nc} = W(\alpha)/W(\mathrm{RN})$. The calculated values were obtained by the power law approximation by Lindhard et al. (1963) which is valid for $E < E_{2c}$.

| Source | $^{210}$Po | | ThC ($^{212}$Bi) | | | ThC' ($^{212}$Po) | $E_{2c}$ |
|---|---|---|---|---|---|---|---|
| Recoil ion | $^{206}$Pb | | $^{208}$Tl | | $^{208}$Pb | | |
| Energy keV | 103 | | 117 | | 168 | | |
| gas | expt | calc | expt | calc | expt | calc | keV |
| $H_2$ | | (0.73) | 0.457[a] | (0.78) | 0.543[a] | (0.93) | 26 |
| $CH_4$ | 0.250[b] | | 0.265[a] | | 0.307[a] | | |
| $C_2H_4$ | 0.236[b] | | 0.269[a] | | 0.321[a] | | |
| $C_3H_6$ | | | 0.272[a] | | 0.281[a] | | |
| $CO_2$ | | | 0.336[a] | | 0.347[a] | | |
| C + 2O | | 0.297 | | 0.316 | | 0.378 | 174 |
| $N_2$ | 0.319[b] | 0.302 | | 0.320 | | 0.384 | 207 |
| Dry air | 0.296[b] | | | | | | |
| 4N + O | | 0.298 | | 0.317 | | 0.379 | 207 |
| $CF_4$ | | | | | | | |
| C + 4F | | 0.280 | | 0.297 | | 0.356 | 174 |
| $CS_2$ | | | | | | | |
| C + 2S | | 0.246 | | 0.262 | | 0.314 | 174 |
| Al (for $CS_2$) | | 0.228 | | 0.242 | | 0.290 | 428 |
| C | | 0.323 | | 0.344 | | 0.411 | 174 |
| O | | 0.284 | | 0.302 | | 0.361 | 241 |
| F | | 0.269 | | 0.286 | | 0.342 | 278 |
| S | | 0.208 | | 0.221 | | 0.264 | 550 |

| Source | $^{218}$Po | | $^{214}$Po | | $E_{2c}$ |
|---|---|---|---|---|---|
| Recoil ion | $^{214}$Pb | | $^{210}$Pb | | |
| Energy keV | 112 | | 147 | | |
| gas | expt | calc | expt | calc | keV |
| $CS_2$ | | | | | |
| C + 2S | | 0.253 | | 0.292 | 174 |
| Al (for $CS_2$) | | 0.237 | | 0.272 | 430 |

[a] Stone and Cochran (1957).

[b] Cano (1968).

**Figure Captions**

Fig. 1. The stopping powers, $S_T$ , $S_n$ , $S_e$, and $LET_{el}$ (= $-d\eta/dx$) for N ions in nitrogen as functions of the energy. The Bragg-like curve ($-d\eta/dR_{PRJ}$) is also shown.

Fig. 2. The Bragg-like curve for N ions in $N_2$. The N ions enter from the right hand side.

Fig. 3. The nuclear quenching factor, $q_{nc}$, calculated for C and S ions in $CS_2$ as a function of energy. The circles are for the measurement and part simulation by Snowden-Ifft (2003). The solid lines are independent element approach and dot-dashed curves are the asymptotic form. The results obtained for C, S and Pb ions using the power law approximation are also shown.

Fig. 4. The values of $q_{nc}$ for C and F ions in $CF_4$ as a function of energy. The results for Pb ions are also shown for 103-168 keV.

Fig. 5. The values of $q_{nc}$ for C and O recoil ions in $CO_2$ as a function of energy. The results for Pb ions are also shown. The $q_{nc}$ values for N ions in $N_2$ are also shown (dot-dash curve).

Fig. 6. The Bragg-like curve, which is closely related to specific ionization, plotted as a function of the projected range $d\eta/dR_{PRJ}$, estimated for 10 to 200 keV recoil C and S ions in $CS_2$. The points are plotted in every 5 keV interval for 10 to 50 keV, in 10 keV interval for 50 to 100 keV, and in 20 keV interval for 100 to 200 keV. The area below each curve (with the unit shown on the axis on the right) expresses the number of ions produced, $N_i$.

Fig. 7. The Bragg-like curve estimated for recoil C and F ions in $CF_4$. The area below each curve expresses the number of ions produced, $N_i$.

Fig. 8. The Bragg-like curve estimated for recoil C and O ions in $CO_2$.

Fig. 9. The Bragg-like curve estimated for recoil Pb ions in $\alpha$-decay in $N_2$, $CO_2$, $CF_4$ and $CS_2$.

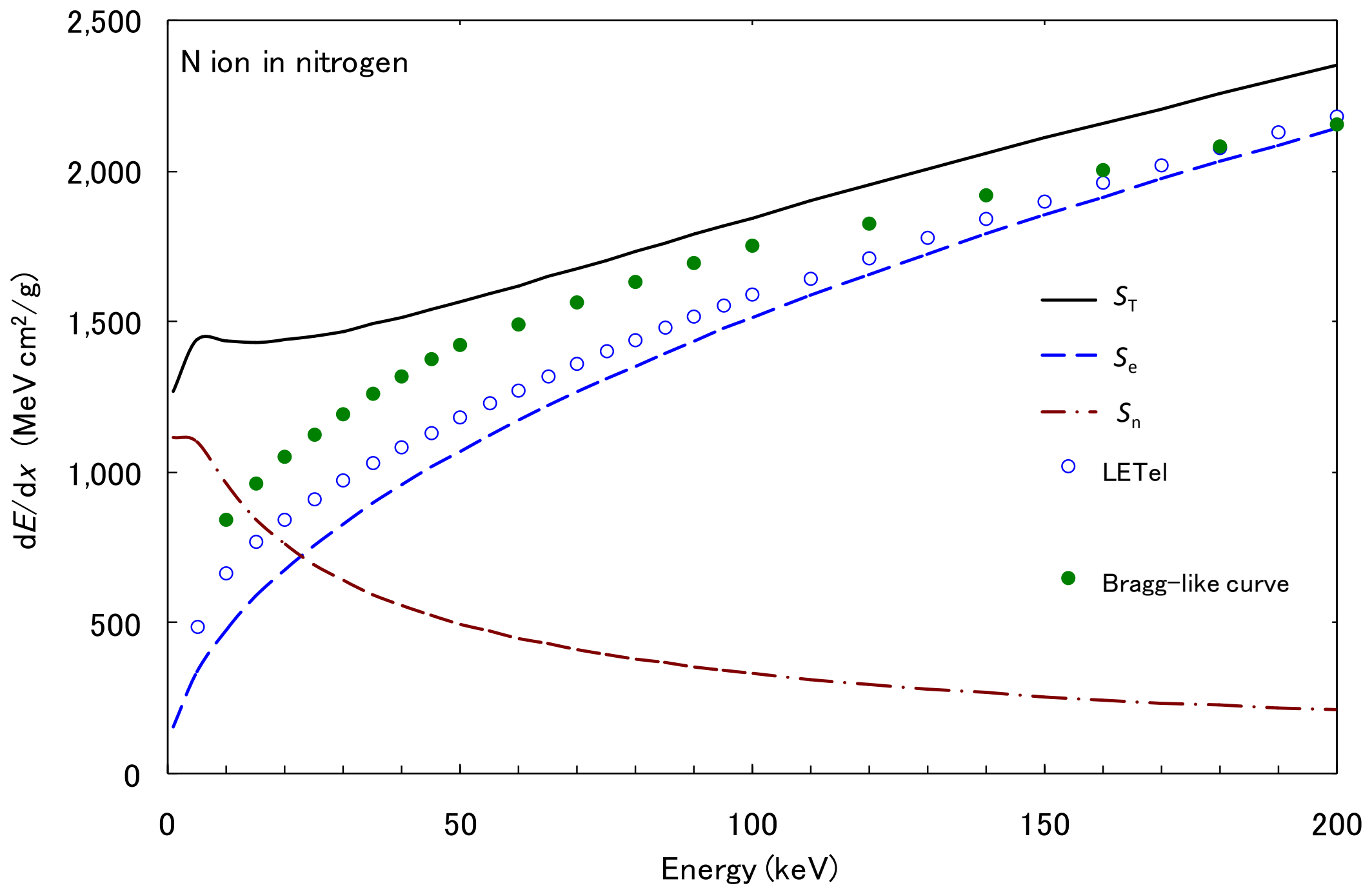

Fig. 1. The stopping powers, $S_T$ , $S_n$ , $S_e$, and LET$_{el}$ (= -dη/d$x$) for N ions in nitrogen as functions of the energy. The Bragg-like curve (-dη/d$R_{PRJ}$) is also shown.

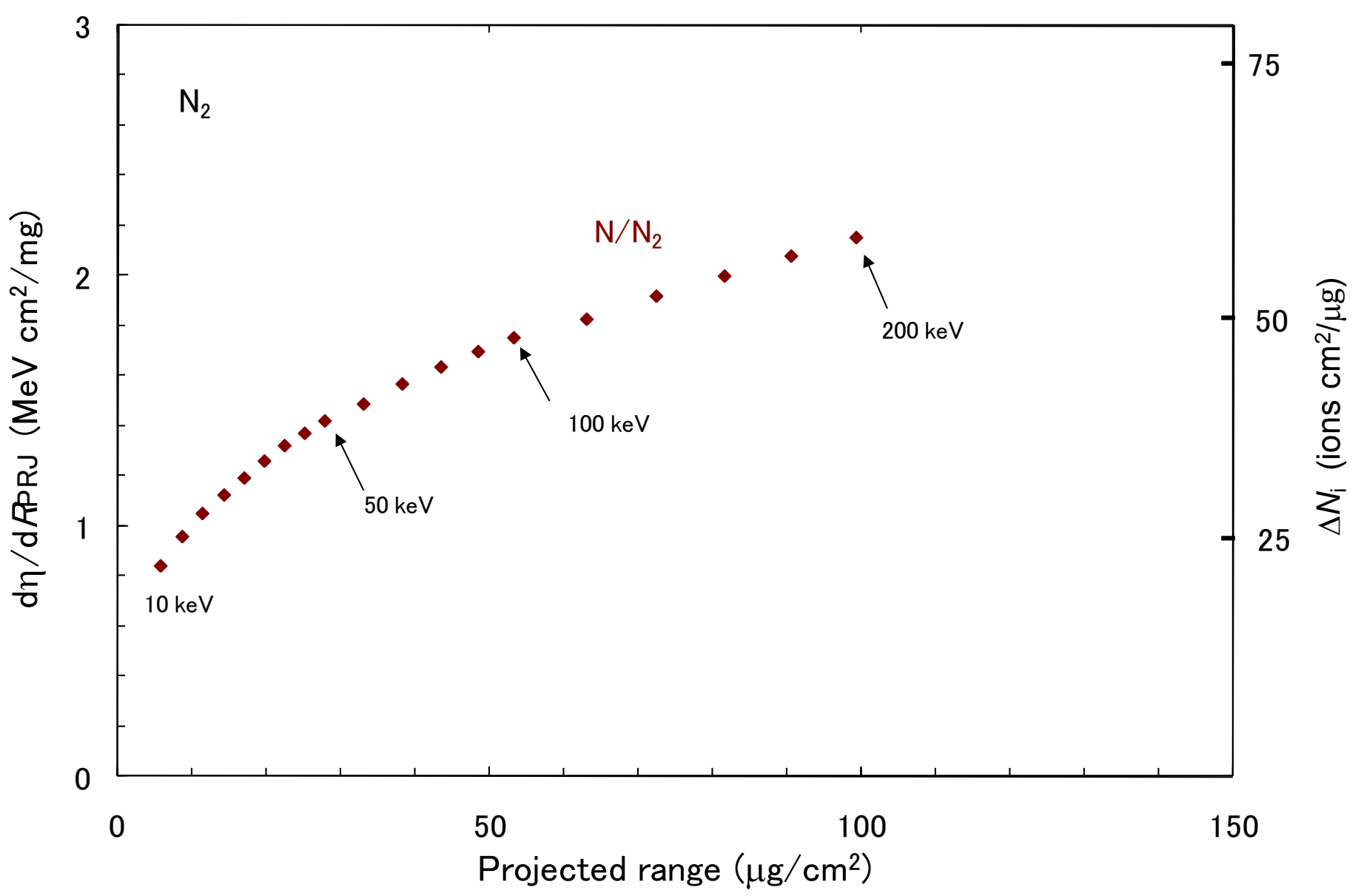

Fig. 2. The Bragg-like curve for N ions in $N_2$. The N ions enter from the right hand side.

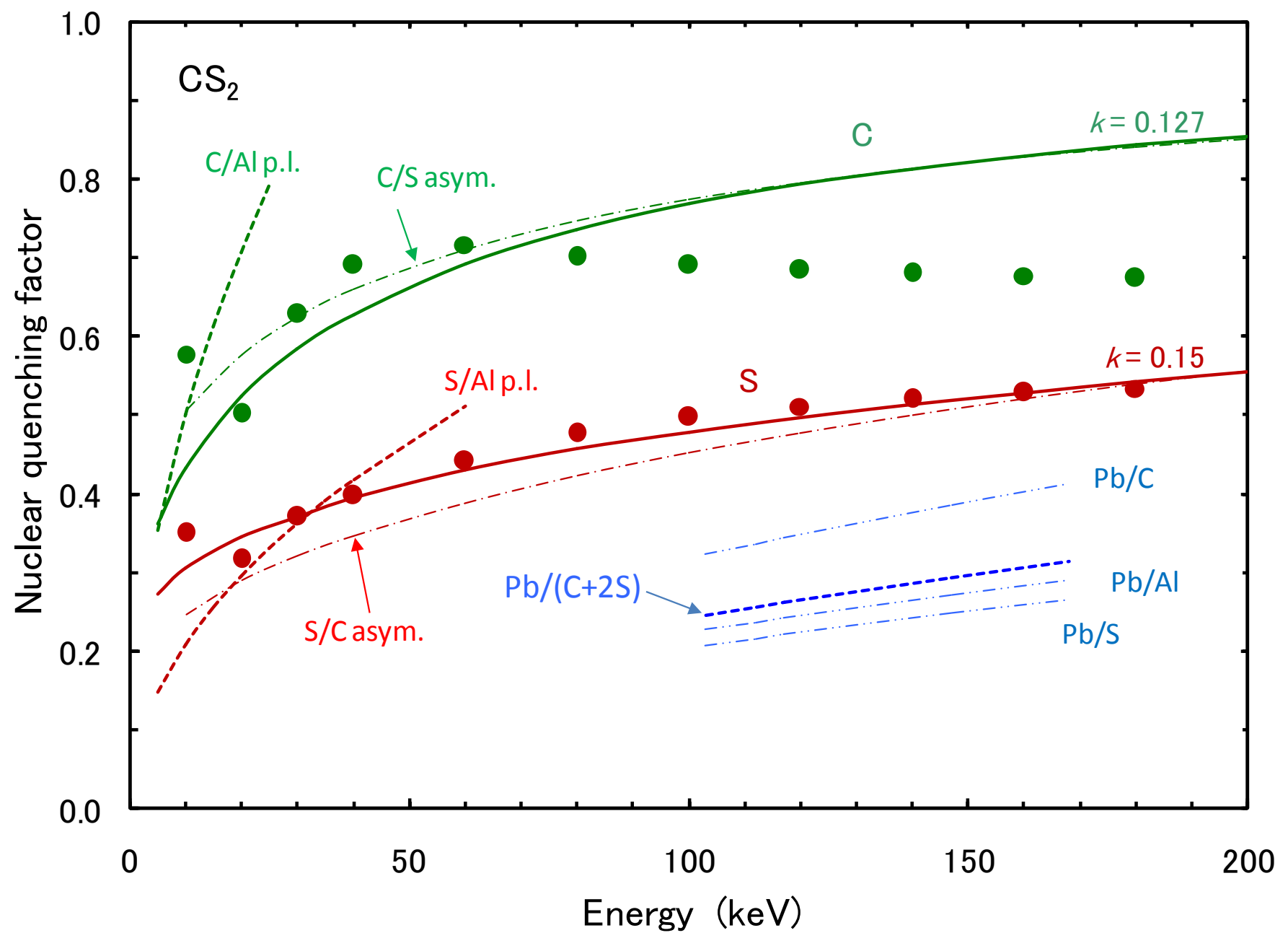

Fig. 3. The nuclear quenching factor, $q_{nc}$, calculated for C and S ions in $CS_2$ as a function of energy. The circles are for the measurement and part simulation by Snowden-Ifft (2003). The solid lines are independent element approach and dot-dashed curves are the asymptotic form. The results obtained for C, S and Pb ions using the power law approximation are also shown.

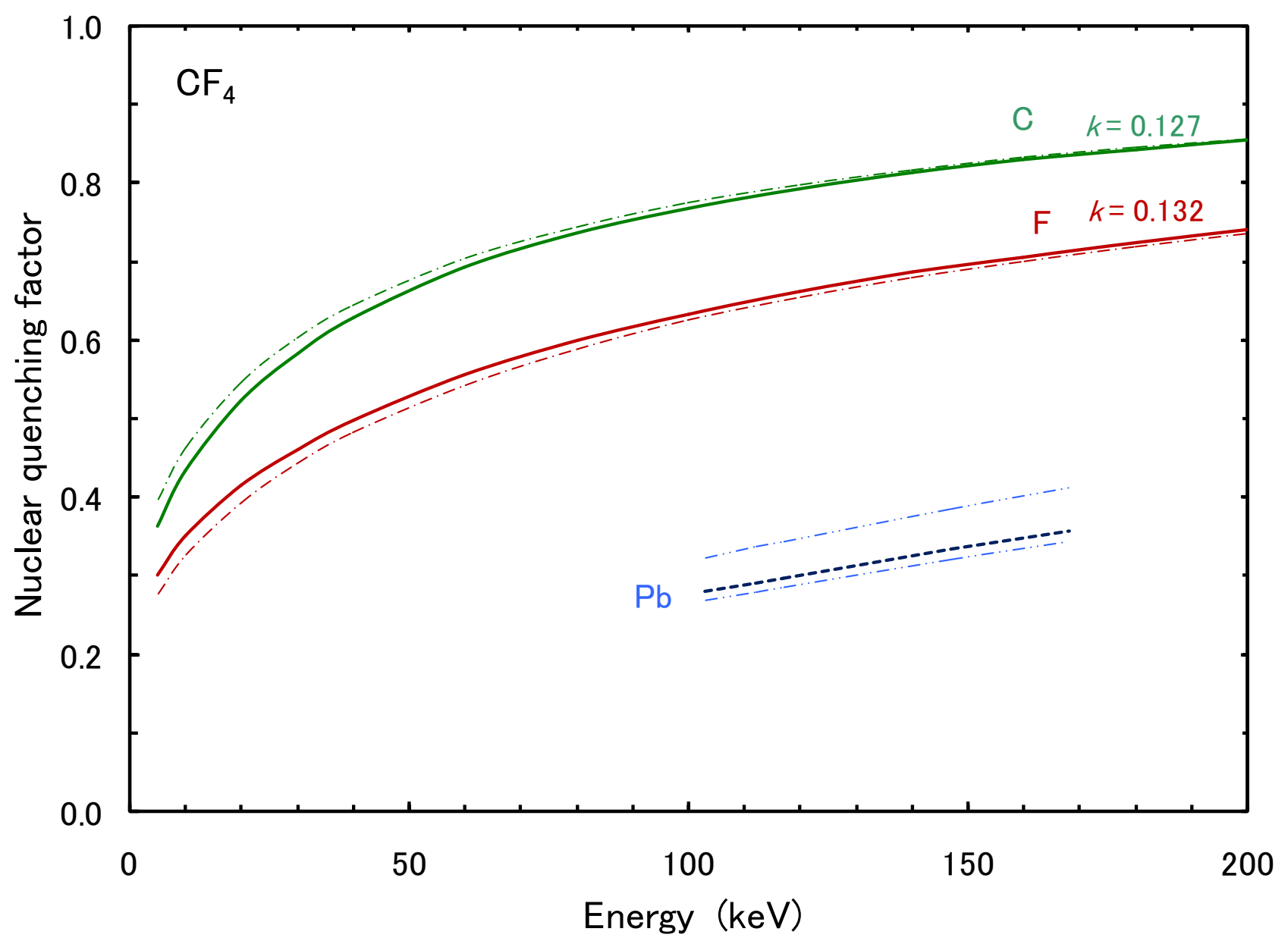

Fig. 4. The values of $q_{nc}$ for C and F ions in $CF_4$ as a function of energy. The results for Pb ions are also shown for 103-168 keV.

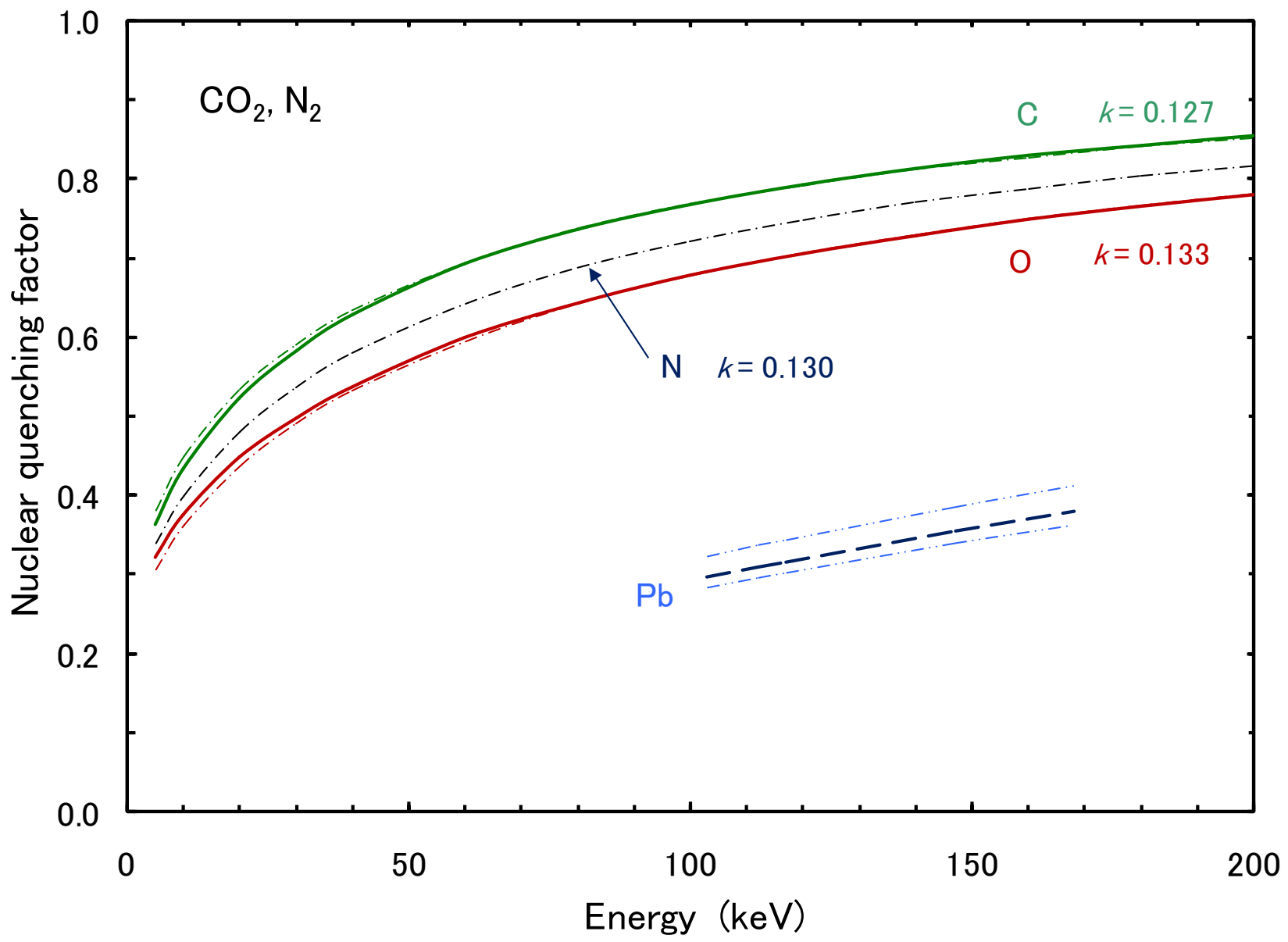

Fig. 5. The values of $q_{nc}$ for C and O recoil ions in $CO_2$ as a function of energy. The results for Pb ions are also shown. The $q_{nc}$ values for N ions in $N_2$ are also shown (dot-dash curve).

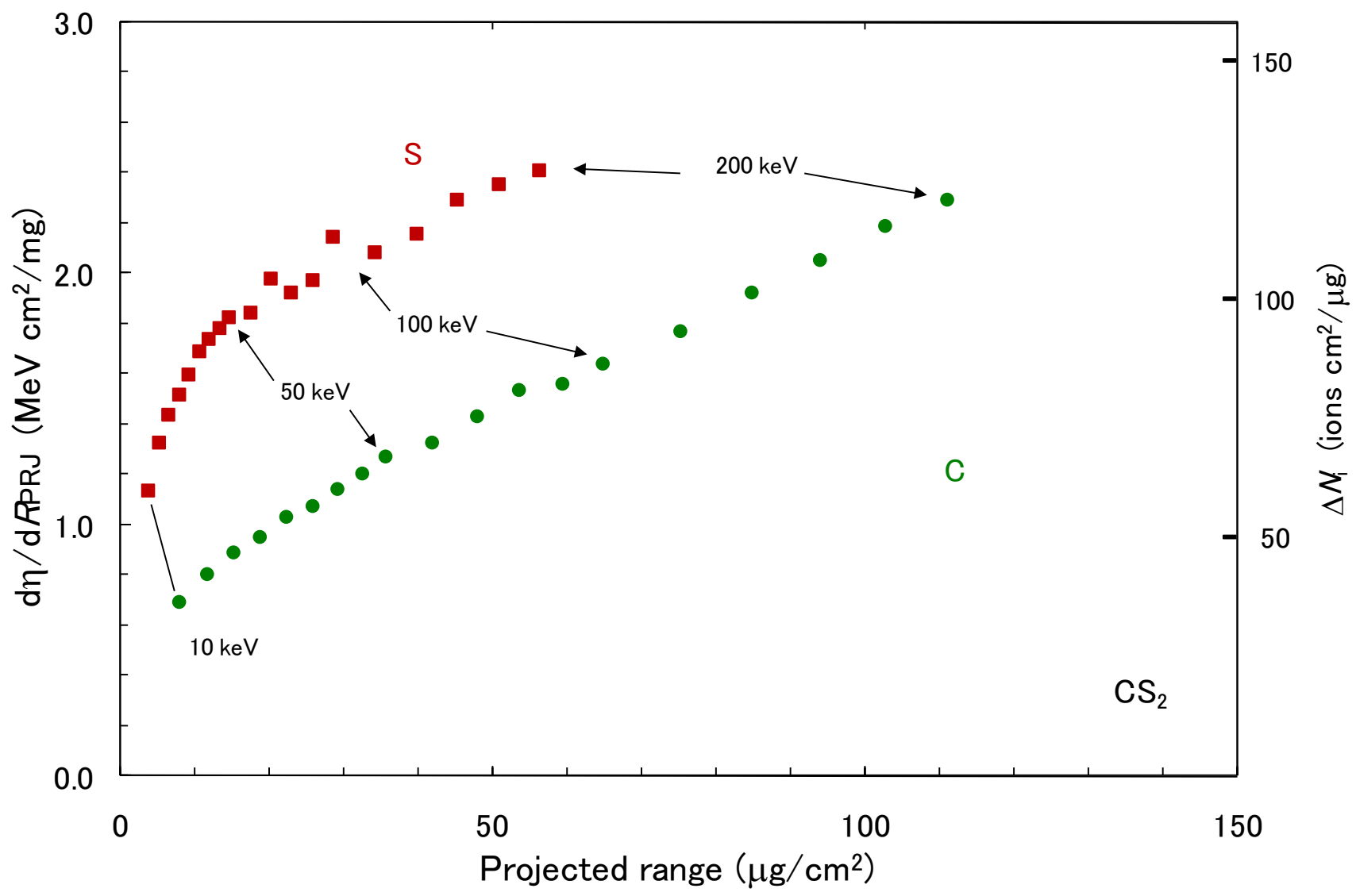

Fig. 6. The Bragg-like curve, which is closely related to specific ionization, plotted as a function of the projected range $d\eta/dR_{PRJ}$, estimated for 10 to 200 keV recoil C and S ions in $CS_2$. The points are plotted in every 5 keV interval for 10 to 50 keV, in 10 keV interval for 50 to 100 keV, and in 20 keV interval for 100 to 200 keV. The area below each curve (with the unit shown on the axis on the right) expresses the number of ions produced, $N_i$.

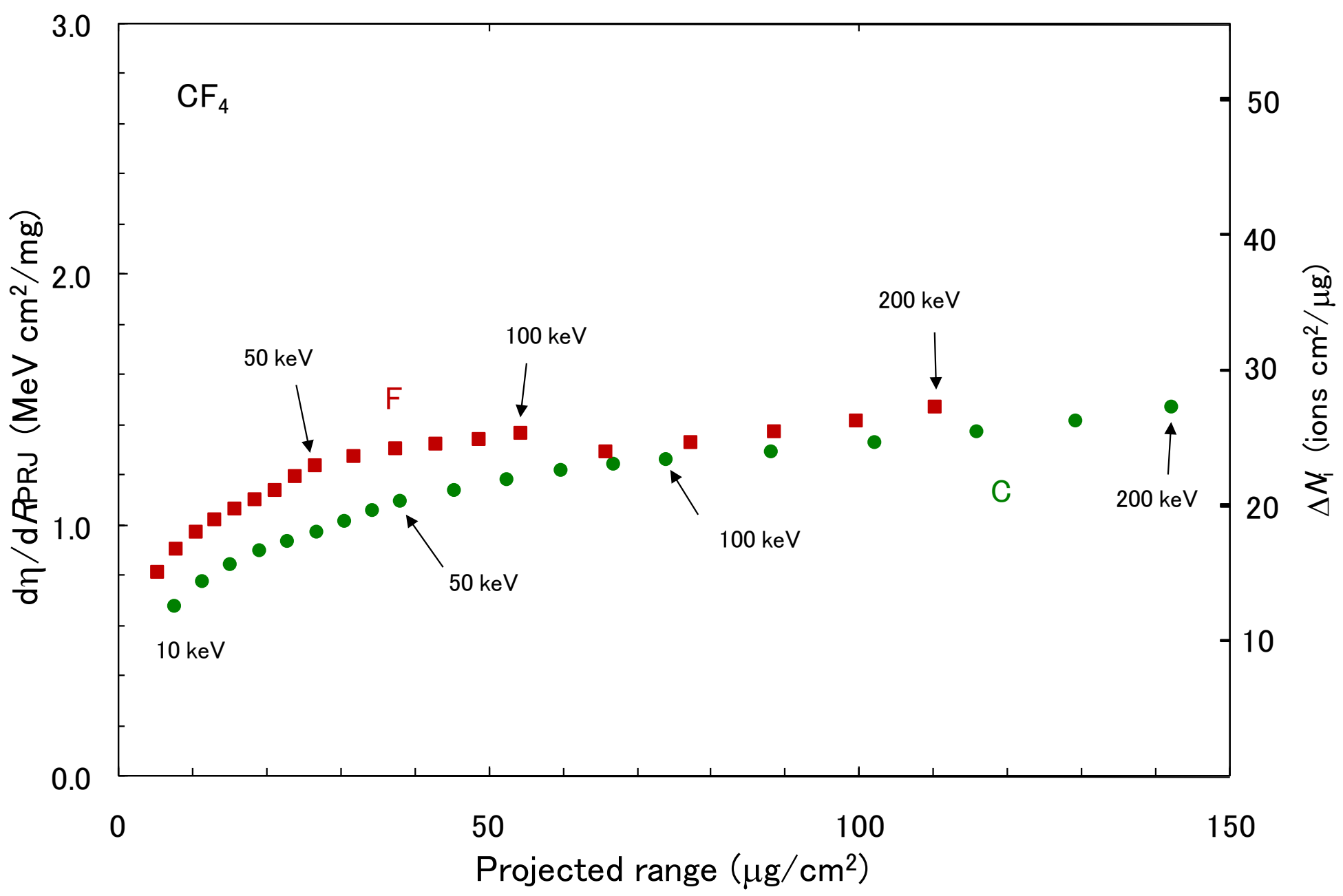

Fig. 7. The Bragg-like curve estimated for recoil C and F ions in $CF_4$. The area below each curve expresses the number of ions produced, $N_i$.

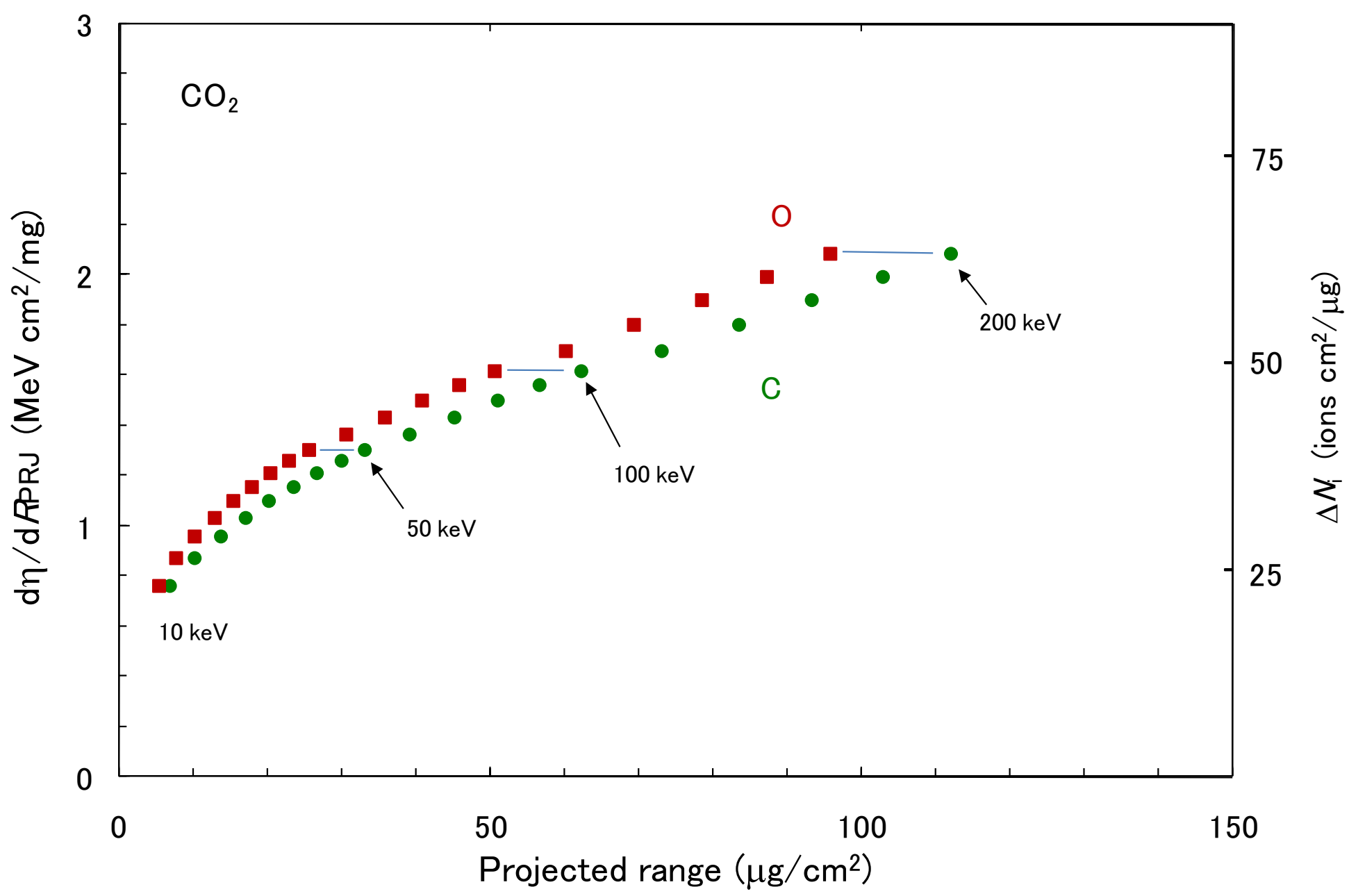

Fig. 8. The Bragg-like curve estimated for recoil C and O ions in $CO_2$.

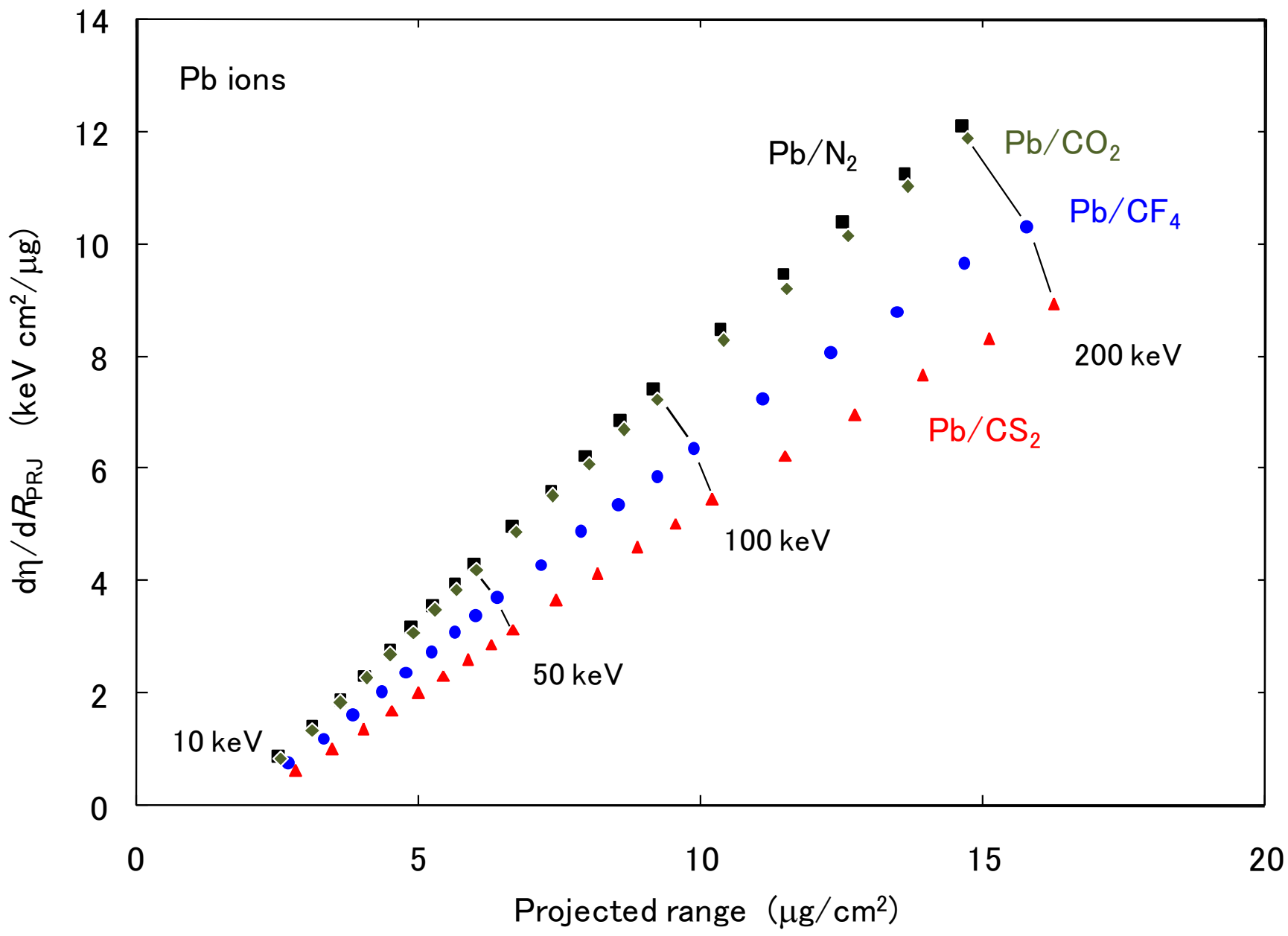

Fig. 9. The Bragg-like curve estimated for recoil Pb ions in α-decay in $N_2$, $CO_2$, $CF_4$ and $CS_2$.